\documentclass[12pt,preprint]{aastex}

\def\alwaysmath#1{\ifmmode{#1}\else{$#1$}\fi}

\def\arcsec{\hbox{$^{\prime\prime}$}}

\newcommand\CHANDRA{{\sl CHANDRA~}}
\newcommand\SYNPHOT{{\sl SYNPHOT~}}
\newcommand\PC{{\sl PC~}}

\def\ltsima{$\; \buildrel < \over \sim \;$}
\def\gtsima{$\; \buildrel > \over \sim \;$}
\def\lsim{\lower.5ex\hbox{\ltsima}}
\def\gsim{\lower.5ex\hbox{\gtsima}}

\shorttitle{47 Tuc} 
\shortauthors{F.R. Ferraro et al.}
\slugcomment{ApJ-in press}

\begin{document}

\title{Blue  Stragglers, Young White Dwarfs and  UV-excess  stars 
in the core of 47~Tuc\footnote{Based on
observations with the NASA/ESA HST, obtained at
the Space Telescope Science Institute, which is operated by AURA, Inc.,
under NASA contract NAS5-26555}} 

\author{Francesco R. Ferraro}
\affil{Osservatorio Astronomico di Bologna, via Ranzani 1,I--40126 Bologna, 
Italy}
\email{ferraro@apache.bo.astro.it}
\author{Nichi D'Amico}
\affil{Osservatorio Astronomico di Bologna, via Ranzani 1,I--40126 Bologna, 
Italy}
\email{damico@bo.astro.it}
\author{Andrea Possenti}
\affil{Osservatorio Astronomico di Bologna, via Ranzani 1,I--40126 Bologna, 
Italy}
\email{phd@tucanae.bo.astro.it}
\author{Roberto P. Mignani}
\affil{ESO, Karl Schwarzschild Str.2, D85748 Garching b. M\"unchen}
\email{rmignani@eso.org}
\author{Barbara Paltrinieri}
\affil{Istituto di Astronomia, Universit\'a La sapienza,Via G.M. Lancisi 29, 
I--00161 Roma, Italy}
\email{barbara@coma.mporzio.astro.it}

\begin{abstract}

We  used a  set of  archived  HST/WFPC2 images  to   probe the stellar
population in  the core of  the nearby Galactic Globular Cluster (GGC)
47 Tuc.   From the  ultraviolet (UV)  Color  Magnitude Diagrams (CMDs)
obtained for $\sim 4\,000$ stars detected  within the Planetary Camera
(PC) field of view we have pinpointed a number of interesting objects:
{\it (i)} 43 blue stragglers stars (BSSs) including 20 new candidates;
{\it (ii)} 12 bright (young) cooling white dwarfs (WDs) at the extreme
blue region of the UV-CMD; {\it (iii)} a large population of UV-excess
(UVE) stars, lying between the  BSS and the  WD sequences.  The colors
of the  WD candidates identified here  define  a clean pattern  in the
CMD,  which  define the  WD   cooling sequence.    Moreover, both  the
location on  the   UV-CMDs and the    number of WDs  are in  excellent
agreement with the theoretical expectations.  The UVE stars discovered
here represent the largest  population of anomalous blue  objects ever
observed in a globular  cluster -- if  the existence  of such a  large
population is   confirmed,  we have   finally found the  long-searched
population of interacting binaries  predicted by the theory.  Finally,
we have investigated the feasibility of  the optical identification of
the    companions of the  binary   X-ray sources  recently detected by
\CHANDRA and of binary millisecond pulsars (MSPs) residing in the core
of  47~Tuc. Unfortunately, the extreme faintness  expected for the MSP
companions together  with the  huge  stellar crowding  in the  cluster
center  prevent statistically reliable  identifications based  only on
positional coincidences. 

\end{abstract}


\section{Introduction} 
\label{sec:intro}

A  large variety of  exotic objects  have been  found to  populate the
cores of the most concentrated Galactic Globular Clusters (GGCs): Blue
Straggler  Stars   (BSSs),  low  mass  X-ray   binaries  (LMXBs),  low
luminosity GC X-ray sources  (LLGCXs), cataclysmic variables (CVs) and
millisecond pulsars (MSPs)  (e.g. see Bailyn 1995).  As  most of these
objects  result from  Interacting Binaries  (IBs), their  formation is
strongly  favoured in  the dense  stellar environment  of a  GC, where
stellar collisions  provide additional channels for  the formation and
evolution of binary systems, e.g., via tidal captures (Fabian, Pringle
\& Rees 1975; Di Stefano  \& Rappaport 1994) and exchange interactions
(Hut,  Murphy  \&  Verbunt  1991).   \\  In  this  framework,  we  are
performing  an  extensive search  for  various  by-products of  binary
evolution in  the cores of a sample  of GGCs selected on  the basis of
their photometric  properties.  In  particular, a large  population of
BSSs (see Ferraro et al.  1997a, 1999; Paltrinieri et al.  1998) and a
few faint ultraviolet (UV) sources suspected to be IBs (see Ferraro et
al.  1997a,  1998, 2000a, 2000b)  have been already discovered  in the
framework of an UV survey of a sample of GGCs carried out with the HST
(see  Ferraro,  Paltrinieri \&  Cacciari  1999  for  a review  of  the
results).   \\ Among  GGCs, 47~Tuc  is definitely  the  most promising
target for this kind of  investigation since it harbours a crowded zoo
of stellar species, including BSSs, LMXBs, candidate CVs and MSPs.  In
particular,  21 BSSs  have been  discovered by  Paresce et  al. (1991,
hereafter P91), and confirmed by De Marchi, Paresce \& Ferraro (1992),
Guhatakurta et al.  (1992) and  Gilliland et al.  (1998). In addition,
20 MSPs have been detected by Camilo et al. (2000), 13 of which are in
binary  systems. Recently,  Freire et  al.  (2001)  presented accurate
timing  position for  most of  them.   Moreover, the  cluster hosts  a
number  of  unidentified  X-ray  sources (Verbunt  \&  Hasinger  1998;
Grindlay et  al. 2001), a dwarf  nova (V2, Paresce \&  De Marchi 1994;
see also Shara et al.  1996) and a suspected interacting binary ($V1$)
found in  the error  circle of  a bright X-ray  source by  Paresce, De
Marchi  \& Ferraro  (1992)- see  also  De Marchi,  Paresce \&  Ferraro
(1993).  Systematic  searches for  variability in the  core of  47 Tuc
have been performed  by many groups (see Shara et  al. 1996; Edmond et
al.  1996;  Albrow et al.   2001), also exploting the  large data-base
provided by the survey of  Gilliland which was originally coinceved to
search for  planetary transits (Gilliland et al 2000).  \\ Here 
 we report on  the results of
recent HST UV  observations of the core of  47 Tuc. These observations
yielded  the  discovery  of  a  large population  of  UV  objects  (\S
\ref{sec:res}) which significantly expand  the list of the interesting
objects (Tables~2, 3, \& 4) known  to reside in the core of 47~Tuc and
surely   will   deserve   a    more   detailed   study   when   higher
resolution/sensitivity  observations  and deeper  imaging/spectroscopy
will be available.  \\ In addition,  we have used the same data-set to
search for the optical counterparts to  LLGCXs and MSPs in the core of
47 Tuc.  The  results of this search and all  the relative caveats are
discussed in section (\S \ref{sec:pos}).
 
\section{Observations and  data analysis}
\label{sec:obs}
  
The data-set  used in the present  work consist of a  series of public
HST/WFPC2 exposures  mapping the core  of the cluster  47~Tuc, taken
through the filters F218W, F336W and F439W.  The observing epochs, the
filter mean wavelength and width, the integration times and the number
of exposures are summaried in Table 1.  The images have been retrieved
from the  public HST Archive after the  on-the-fly recalibration with
the most recent  reference files and tables. Images  taken throught the
same filter have been combined  and cleaned from cosmic ray hits using
the automatic WFPC2 {\it  Archive Association Procedure}.  In all the
selected observations, the core of  47 Tuc was roughly centered on the
Planetary Camera  (PC) chip, which has the  highest angular resolution
($0\farcs045$/px). We  thus decided  to focus our  analysis on  the PC
frames  only.   Note  that the  F336W  images  have  been taken  at  a
different epoch and with a different telescope roll angle with respect
to the F218W/F439W  data-sets. For this reason, they  cover only $\sim
70\%$ of  the \PC field of view  imaged in the other  two filters.  The
standard procedure  described in Ferraro  et al. (1997a)  was adopted.
In short,  in order to perform a  deep search for faint  UV objects we
used the combined F218W image  (with an equivalent total exposure time
of 3200  sec) as  the reference frame  for object detection.  The same
image was  also used as a  reference for the  frame registration.  The
photometric reductions have been  carried out using ROMAFOT (Buonanno
et  al. 1983), a  package specifically  developed to  perform accurate
photometry in crowded fields.  The  object list derived from the F218W
image was  then matched  to the corresponding  ones obtained  from the
F336W  and F439W  images.   A final  catalogue  with the  instrumental
magnitude in each filter and coordinates has been compiled for all the
identified   stars.    The   instrumental  magnitudes   were   finally
transformed to the STMAG  photometric system using Table~9 by Holtzman
et al. (1995).
 
\section{Results and Discussion} 
\label{sec:res}

\begin{figure}
\plotone{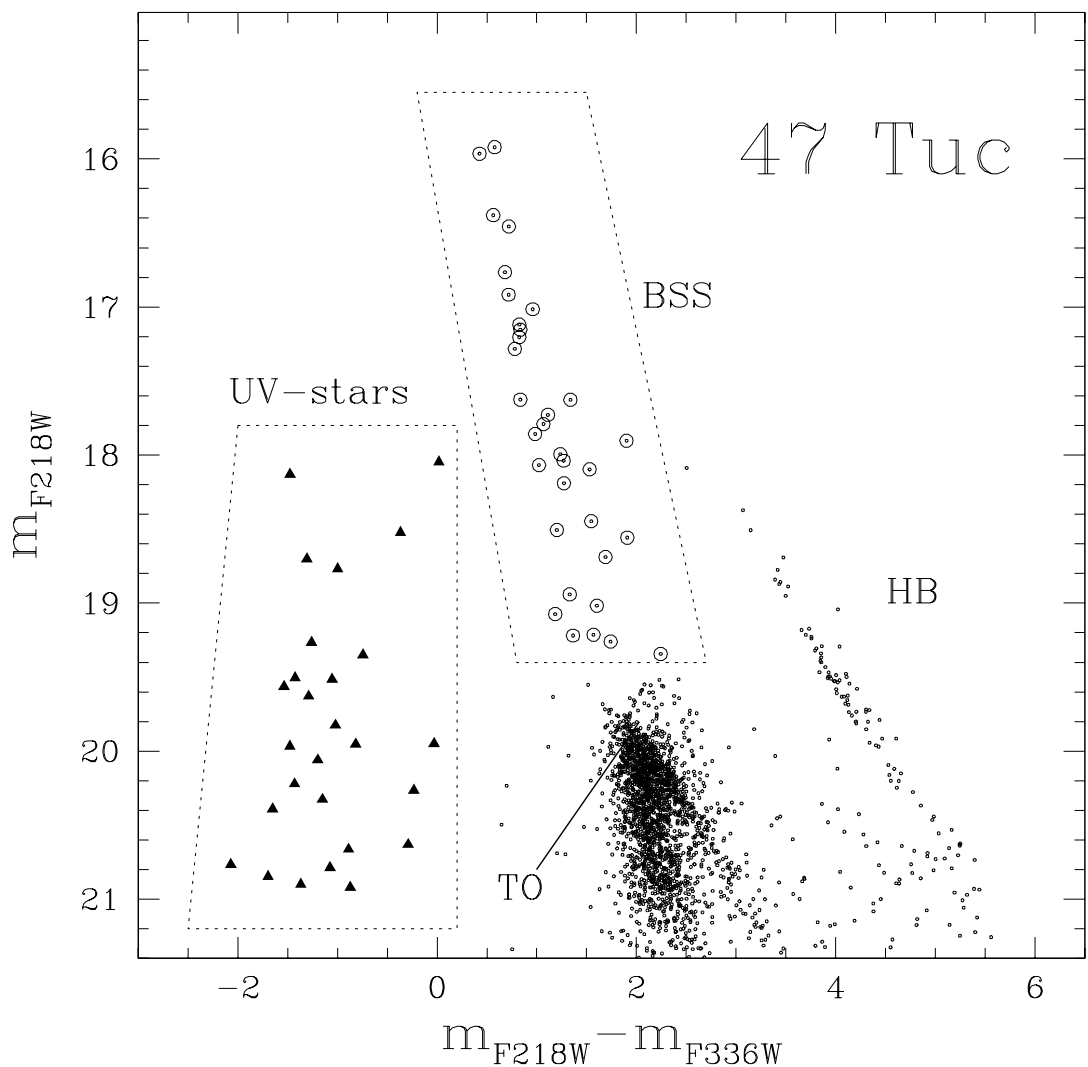} 
\caption{$(m_{F218W}-m_{F336W},  m_{F218W})$  CMD  obtained  from  the
photometry of more than 4\,000  stars detected in the Planetary Camera
(PC) field of  view. BSSs (see Table 2) are  highlighted by large open
circles  while  the  population  of  UV-stars  is  plotted  as  filled
triangles.       The     main      evolutionary      sequences     are
labelled. \label{fig:map336}}
\end{figure}

Figure~\ref{fig:map336}  shows the  $(m_{F218W}, m_{F218W}-m_{F336W})$
Color  Magnitude Diagram (CMD)  obtained from  the photometry  of more
than 4\,000 stars identified in the \PC field of view 
of the combined F218W image.  As it can be seen, 
the overall morphologies of the branches
in  this   diagram  are  deformed   with  respect  to   the  classical
$(V,B-V)$-plane.   For clarity,  the main  evolutionary  sequences are
labelled in the diagram (see  also Figure 1 in Ferraro, Paltrinieri \&
Cacciari   1999).   Figure~\ref{fig:map439}  shows   the  $(m_{F218W},
m_{F218W}-m_{F439W})$  CMD,  where  the  F439W  detections  have  been
included. \\ As expected, the cluster emission at these wavelengths is
dominated by hot  stars.  In particular, since 47~Tuc  is a metal rich
cluster with  essentially a red  Horizontal Branch (HB), hot  HB stars
are  missing, and  the brightest  population turns  to be  composed of
BSSs.     Schematically,   two    prominent    features   appear    in
Figure~\ref{fig:map336}  and  are  fully   confirmed  in  the  CMD  of
Figure~\ref{fig:map439}:  (1) the population  of BSSs  (highlighted by
open circles) which  defines a clean sequence spanning  $\sim 3.5$ mag
and is reminiscent of the  sequences already observed in other UV-CMDs
like, e.g., M3  (Ferraro et al. 1997) and M80  (Ferraro et al.  1999);
(2) the large  number of UV-excess stars (marked  by filled triangles)
which   populate   the   extreme    blue   region   of   the   diagram
($m_{F218W}-m_{F336W}<0$).   Although the observations  presented here
span three different epochs,  an independent search for variability is
beyond the aims  of this paper and we refer to  the extensive work by
Shara et al.  (1996), Edmonds et al. (1996,  hereafter E96) and Albrow
at al. (2001) in order to  check for the variability of the UV objects
found in this work. \\ In  the following, we will describe and discuss
separetely the results of our analysis of these peculiar populations.

\subsection{Blue Straggler Stars}
\label{subsec:bss}

The presence  of BSSs in the  core of 47  Tuc is well known  since the
pioneering  UV observations  performed  with the  Faint Object  Camera
(FOC) by P91,  who discovered 21 BSSs.  Most  of these identifications
are  now confirmed by  the CMDs  shown in  Figure~\ref{fig:map336} and
Figure~\ref{fig:map439},  which   unveil  a  total  of   43  BSS  with
$m_{F218W}<19.4$.   Magnitudes  and positions  for  these objects  are
listed  in  Table~2.   All  the  coordinates are  in  \PC  pixel  units
($0\farcs045$/px) and have been  measured in the F218W image reference
frame.  In  Table 2  we adopted the  BSS identification code  (BSS \#)
originally used by  P91 and we followed this  numeration for the other
candidates  successively discovered  by Guhathakurtha  et  al.  (1992,
hereafter  G92)  and  E96,  as  well  as for  the  additional  20  new
candidates   discovered   in   the   present   work.    The   original
identification number by G92 is also listed in column 8, together with
the one used by E96 to flag variables objects (namely V1, V3, V6, V10,
V11,  V12).  Some  of these  stars have  been successively  studied by
Gilliland  et  al.   (1998).   In  particular, V10  (our  BSS-24)  was
classified as eclipsing  variables by E96, while V12  (our BSS-16) was
suggested to  be a low amplitude  SX Phe variable in  a binary system.
On the  basis of  its light curve  shape, V11\footnote{a.k.a  AKO9, as
named  in the  original  work by  Auriere,  Koch-Miramond \&  Ortolani
(1989)}  (our BSS-26)  was classified  by  E96 as  an Algol  eclipsing
binary, with  a period of $\sim 1.1$  day.  The true nature  of V11 is
not  completely understood.  Indeed, while  the CMD  presented  by E96
shows this object as ``located  near the main sequence turn-off'', our
photometry shows that it is in  the BSS region. In addition, its light
curve  is known  to  show  anomalous features  like,  e.g., the  rapid
brightening  detected by  Minniti  et al.  (1997).\\  Only three  BSSs
discovered  by  P91 (namely  BSS9,  BSS11  and  BSS17) have  not  been
confirmed as BSS  by our criteria. In particular,  BSS11 is located in
the region between  the cluster Turn Off (TO) and  the BSS sequence in
Figure~\ref{fig:map439}, but  it is  below the threshold  assumed here
($m_{F218W}=19.4$).  BSS9  (ID \#2800  in our list)  is the  object at
$m_{F218W}\sim   18.1$  and  colour   $(m_{F218W}-m_{F439W})\sim  3.3$
located between the BSS sequence and  the HB.  BSS17 (ID \#1798 in our
list)  has  been  classified  as  UV-excess star  (UVE2,  see  Section
\ref{subsec:uve})  since  its  location  in  the  CMD  ($m_{F218W}\sim
18.05$; $(m_{F218W}-m_{F439W})\sim  0.9$) is significantly  bluer than
that of the BSS sequence.

\subsection{Young White Dwarfs}
\label{subsec:wd}
 
\begin{figure}
\plotone{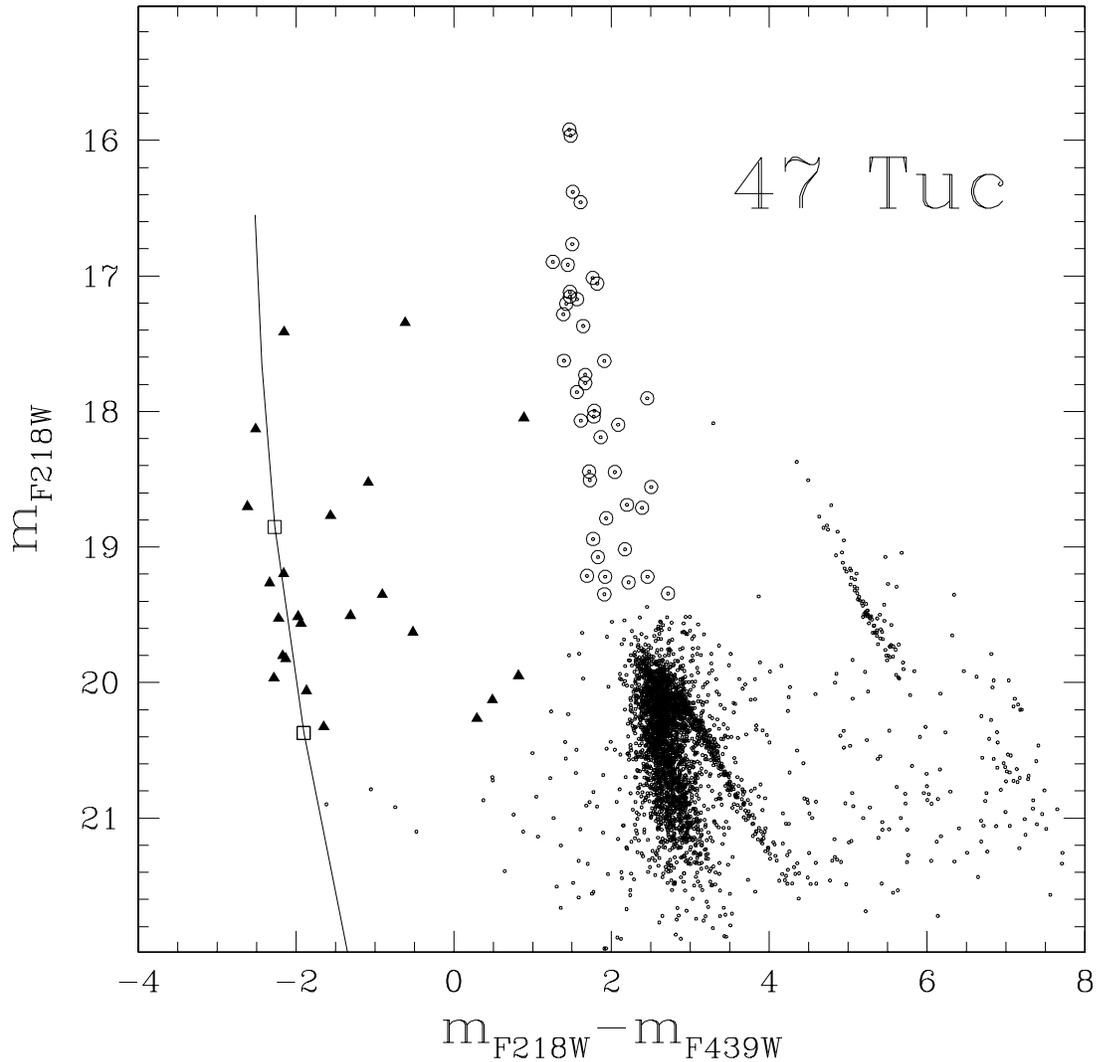}
\caption{($m_{F218W},  m_{F218W}-m_{F439W}$) CMD  for  the same  stars
shown in Figure~\ref{fig:map336}.  As in Figure~\ref{fig:map336}, BSSs
(see Table 2) are highlighted by large open circles and the population
of UV-stars have been plotted as filled triangles.  The theoretical WD
cooling  sequence, computed  using the  Wood (1995)  models,  has been
overplotted to  the data.  The two  open squares along  the WD cooling
sequence mark the location of 3 and 13 million year old cooling WDs.
\label{fig:map439}}
\end{figure}

The   extended    colour   baseline    provided   by   the    CMD   in
Figure~\ref{fig:map439} allows us to further investigate the nature of
the  UV-excess  stars,   identified  in  Figure~\ref{fig:map336}.   In
particular,  we  note  how  the  bluest side  of  the  UV-excess  star
distribution seems to  define a sort of sequence.  This would suggest
that at  least part of the  UV-excess stars could be  white dwarfs, accordingly with the previous findings by (Paresce, De Marchi \& Jedrzejewski 1995). To
check this hypothesis we have computed the expected location of the WD
cooling sequence in the $(m_{F218W},m_{F218W}-m_{F439W})$ CMD.  To do
this,  we run  the theoretical  cooling sequence  for  $0.5 M_{\odot}$
hydrogen-rich WDs  (Wood 1995)  through the standard  IRAF simulation
package \SYNPHOT in order to  derive the UV-colours and magnitudes for
the  theoretical WD  sequence.  A  reddening of  $E(B-V)=0.04$  in the
direction  of  47  Tuc  was  assumed.  The  computed  WD  sequence  is
overplotted  to the  data in  Figure~\ref{fig:map439}.  As  it  can be
seen,  the  cooling  sequence  nicely  fits the  bluest  side  of  the
UV-excess distribution  and fully  supports the conclusion  that about
half of the UV-excess stars detected  in the core of 47~Tuc are indeed
WDs.  In addition, by comparing  the observed WD distribution with the
expected location of 3 and 13  million year old cooling WDs (marked by
the   two   open   squares   along   the  WD   cooling   sequence   in
Figure~\ref{fig:map439}) we  can conclude  that   the  WD population
discovered here  basically consists of  very young objects.  \\ Though
the     sample     presented     in    Figure~\ref{fig:map336}     and
Figure~\ref{fig:map439} is surely  affected by incompleteness at least
at the  faint end  of the sequence  ($m_{F218W}>20.5$), we can  try to
perform  a {\it  first  order} test  to  check if  the  number of  the
candidate bright  WDs observed in the  central region of  47~Tuc is in
agreement with the theoretical  expectations.  To do this, we computed
(accordingly with the suggestion by  Richer et al. 1997) the number of
WD expected  in the \PC field  of view by comparison  with the observed
number of a tracer population with known lifetime (as the HB).  Taking
into account  that in the  field of view  of the \PC we  observed $\sim
100$ HB stars, and assuming  the HB lifetime $t_{HB}=10^8 yr$ (Renzini
\&  Fusi Pecci  1988), we  derive from  their eq.   (1)  the following
relation for our sample:

\begin{equation}
N_{WD} (<m^{*}_{F218W}) = 10 ^{-6}~~t_{cool} ( <m^{*}_{F218W})
\end{equation}

Here  $N_{WD} (<m^{*}_{F218W})$  is the  number of  WDs  brighter than
magnitude   $m^{*}_{F218W}$  and  $t_{cool}(<m^{*}_{F218W})$   is  the
corresponding  cooling time in  yrs.  According  to the  Wood's models
(Wood  1995), $t_{cool}  (m_{F218W}<18.9)\sim  3 \times  10^6$ yr  and
$t_{cool}  (m_{F218W}<20.4)\sim 1.3 \times10^7$  yr. Thus,  Eq. (1)
predicts  $N_{WD}(m_{F218W}<18.9) \sim 3$  and $N_{WD}(m_{F218W}<20.4)
\sim 13.$  Since from  Figure~\ref{fig:map439} we count  3 and  12 WDs
brighter  than  $m_{F218W}= 18.9$  and  $20.4$,  respectively, we  can
conclude that the agreement  with the theoretical prediction turns out
to be exceptionally good.  Thus, we can reasonably assume that all the
objects  with colour $m_{F218W}-m_{F439W}<-1.6$  are young  cooling WD
candidates. \\  Magnitudes, colours and positions for  these stars are
listed in Table~2.

\subsection{UV-Excess stars}
\label{subsec:uve}
 
\begin{figure}
\plotone{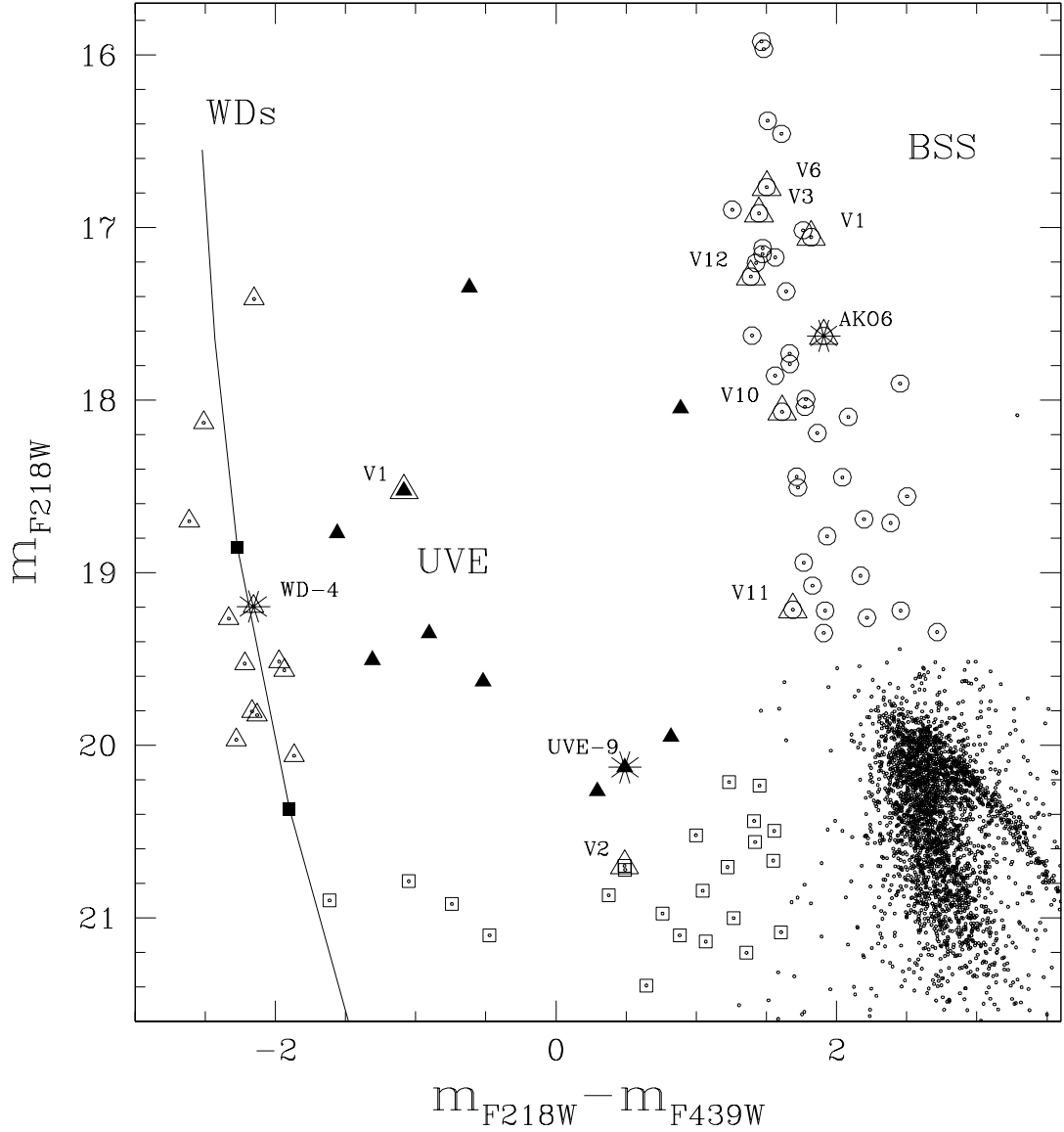}      
\caption{Zoom   of  the   ($m_{F218W},   m_{F218W}-m_{F439W}$)-CMD  of
Figure~\ref{fig:map439} showing in more  detail the nature of the blue
populations in the core of 47~Tuc.  BSSs (see Table 2) are highlighted
by  large open circles,  with variables  further highlighted  by large
empty triangles  and labelled with  their names.  WDs  candidates (see
Table 3)  are plotted as  open triangles.  The theoretical  WD cooling
sequence from Wood (1995) models  has been overplotted to the data (as
in  Figure~\ref{fig:map439}), with the  location of  3 and  13 million
year old cooling  WDs now marked by filled  squares.  Bright UVE stars
(see Table 4) are plotted as filled triangles. The presence of a large
population of {\it  faint} UVE stars (empty squares)  is also shown in
the Figure.  Peculiar objects are higlighted by a larger open triangle
and labelled with  their names.  The three blue  objects (namely WD-4,
BSS-2(AKO6) and UVE-9) which have been found to be close to the three
binary  MSPs -  see Table~5  - are  marked with  a large  asterisk and
labelled with their names.
\label{fig:mapzoom}}
\end{figure}

Although about  half of  the UV-excess stars  detected in the  core of
47~Tuc turn out to be young  cooling WDs, there is still a significant
number  of objects  in Figure~\ref{fig:map439}  lying between  the BSS
sequence  and the  WD cooling  sequence.  A  zoom of  the ($m_{F218W},
m_{F218W}-m_{F439W}$) CMD is  shown in Figure~\ref{fig:mapzoom}, where
we  have now  marked  with  large empty  triangles  the WD  candidates
identified in the  previous section and listed in Table  3.  As it can
be seen from  Figure~\ref{fig:mapzoom}, there are at least  a dozen of
UV-excess stars (plotted as filled triangles) lying between the WD and
the BSS  sequence.  In  the following, we  refere to these  objects as
UV-excess  stars  (hereafter  UVE).   \\ The  identification  numbers,
magnitudes and  positions for the 11  brightest ($m_{F218W}<20.5$) UVE
stars are listed  in Table~4. UVE-3 has been  identified with the blue
object V1 (Paresce, De Marchi \& Ferraro 1992, hereafter PDF92), which
is  now clearly confirmed  to be  an object  with strong  UV emission.
UVE-2 was originally classified as a BSS (BSS17) by P91. However, this
object is bluer  than the BSS sequence, an  evidence also confirmed by
previous CMDs  of the field (see  Figure~5 by  De Marchi,
Paresce  \& Ferraro  (1993), where  it clearly  stands out from  
the BSS sequence).
 UVE-2  is also the  variable PC1-V36 discussed by  Albrow et
al. (2001)  and classified  as a semi-detached  binary.\\ In  a recent
review, Ferraro et  al.  (2000a) listed 10 UVE  objects (associated to
LLGCXs) detected  in the  core of 6  GGCs.  The  population discovered
here  is much  larger  than that  previously  found in  any GGCs  (see
Paresce et al. 1991; Paresce \& De Marchi 1994; Shara \& Drissen 1995;
Cool  et  al.   1995,  1998;  Ferraro  et  al.  1997,  2000a,  2000b).
Moreover, although  we conservatively  listed in Table  4 only  the 11
brightest UVE sources, a  population of {\it faint} ($m_{F218W}>20.5$)
UVE  candidates  (marked  as  open  squares)  is  clearly  visible  in
Figure~\ref{fig:mapzoom}, with  more than 30 UVE  sources counted down
to  $m_{F218W}\sim  21.3$.   Unfortunately,  the  faintness  of  these
objects in  all the  images prevents to  firmly assess  their colours.
Among the {\it faint} UVE stars,  we have identified the dwarf nova V2
discovered by  Paresce \&  De Marchi (1994)  and the blue  variable V3
(suspected to be a CV) discovered by Shara et al (1996).  V2 ($\#5400$
in our list) appears in a quiescence state at the epoch covered by the
present  obervations,  as  it  is quite  faint  ($m_{F218W}=20.7$  and
$m_{F439W}=20.2$)  in  both the  F218W  and  F439W  images.  Also  the
variable V3 ($\#713$ in our list)  appears to be faint at the epoch of
the observations: it  is one of the faintest objects  in the UV region
of  the CMD  shown in  Figure~1 and  it is  not visible  in  the F439W
images.\\  From the  number  of bona-fide  identifications and  likely
candidates pinpointed by our  UV-CMDs, we conservatively conclude that
{\it  there  is  a clear  indication  for  the  existence of  a  large
population of faint UVE stars in the core of 47 Tuc}.  This conclusion
is supported  by additional, independent, observations.  While we were
writing this  paper, a work  by Knigge et  al. (2001) reported  on the
possible detection of  a large population of UVE  stars ($\sim 30$) in
the core of 47~Tuc  from far-ultraviolet photometric and spectroscopic
HST/STIS observations.   The availability of  a spectroscopic analysis
suggested that  these stars (or at  least part of them)  could be CVs.
If this is  the case, the UVE objects  discovered here could represent
the  brightest  extension  of  the  long searched  population  of  CVs
predicted by the  models (see Di Stefano \&  Rappaport 1994) and never
found by previous  observations.  \\ However, alternative evolutionary
scenarios exist, which can  produce objects populating the same region
of the  CMD.  For  instance, Edmonds et  al. (1999) suggested  that at
least one UVE  star found in the center of NGC6397  (Cool et al. 1998)
is actually a low-mass helium  WD, resulting from the stripped core of
a red giant (Castellani \& Castellani 1993).
 
\section{Absolute positions: limits to the search of optical counterparts
to X-ray sources and MSP companions}
\label{sec:pos}

Most of the exotic objects (BSSs, LMXBs, CVs, MSPs) hosted in the core
of GGCs are  thought to result from the evolution  of various kinds of
binary systems. In particular, when a binary system contains a compact
object  (like a  neutron star  or a  white dwarf)  and a  close enough
secondary,  mass  transfer  can  take  place, producing  an  IB.   The
streaming gas,  its impact on the  compact object, the  presence of an
accretion disk  can give  such systems observational  signatures which
make them  stand out above  ordinary cluster stars.   These signatures
include X-ray emission, significant radiation in the ultraviolet (UV),
emission lines, or  rapid time variations. For this  reason, the study
of   peculiars   objects  usually   benefits   from  a   comprehensive
multiwavelength approach.   \\ In the case  of the UVE  stars found in
other GGCs, some were found  coincident with X-ray sources (Ferraro et
al. 1997b,  2000a) while others, like  the ones discovered  by Cool et
al. (1995) in NGC6397, by Carson et al.  (1999) in $\omega$-Cen and by
Ferraro et al.  (2000b) in the  core of NGC6712, were found to exhibit
significant  H$\alpha$ emission.   These  multiwavelength observations
allow  one to  classify  UVE stars  as  high-confidence IB  candidates
and/or probable CVs.  For  this reason, it is particularly interesting
here to investigate possible associations between the blue populations
(BSSs,  WDs,  and UVE  stars),  with some  of  X-ray  sources and  MSP
companions which are known to harbour  in the core of 47~Tuc. \\ Since
the first  guess to  claim an optical  identification is based  on the
positional  coincidence, $X$  and $Y$-pixel  positions of  the objects
detected in  the \PC have  been converted to  $RA$ and $DEC$  using the
task METRIC in the IRAF/STSDAS package, which also accounts for the
geometric distorsions of the PC.  Of course, the accuracy on the final
coordinates  is  entirely dominated  by  the  intrinsic  error on  the
absolute coordinates of  the GSC1.1 ({\it Guide Star  Catalog }) stars
(Lasker et al. 1990),  which are used  both to  point HST  and to  compute the
astrometric  solution in the  focal plane.   According to  the current
estimates, the  mean uncertainty on  the absolute positions  quoted in
the    GSC1.1   is   $\approx    1\farcs5$   ($3-\sigma$    level   of
confidence). Thus,  we conservatively assume $\sim 2''$  as the global
error on  the absolute  positions of our  objects.  \\ Note  that such
estimate  is  consistent  with  the  average  difference  between  the
absolute  positions  of  the  same   set  of  stars,  as  measured  in
independent HST  pointings.  For instance, G92  measured the absolute
position  of its star  {\it E}  to be  $ RA(2000)_{G92}=  00^{\rm h}\,
24^{\rm m}\,  05\fs 33$ and  $DEC(2000)_{G92}=-72\arcdeg\, 04\arcmin\,
54\farcs 46$, while star ($\# 2187$ in our catalog) turns out to be at
$   RA(2000)=    00^{\rm   h}\,    24^{\rm   m}\,   04\fs    92$   and
$DEC(2000)=-72\arcdeg\, 04\arcmin\, 54\farcs  25$, i.e.  at $2''$ from
the G92 determination, which  fully confirms the uncertainty estimated
above.

\subsection{X-ray sources}
\label{subsec:xray}
 
 Verbunt \&  Hasinger (1998, hereafter VH98)  have recently re-analized
 all the ROSAT/HRI  images of 47~Tuc obtained in  the recent years and
 finally produced  a list  of accurate positions  and fluxes  for nine
 X-ray sources detected in the cluster core.
However,  while we  were writing  this paper  Grindlay et  al.  (2001,
hereafter G01)  reported the discovery of additional  $\sim$ 100 X-ray
sources detected from \CHANDRA  observations. This catalog, listing 108
X-ray  sources  with  $L_X<10^{30.5}$  erg  $s^{-1}$,  represents  the
largest and  most complete X-ray  source catalog ever published  in 47
Tuc. For this  reason, in the following, we will  adopt Table~1 of G01
as reference list for the positions  of X-ray sources in 47 Tuc. Among
these, 36 \CHANDRA sources fall within  the field of view of the PC. \\
In order account for RA/DEC off-sets between the HST coordinate system
and the \CHANDRA one, we have  converted the coordinates listed in Table
1 of G01 in \PC pixels using the WFPC2 astrometric solution and we have
shifted the  G01 catalog to match  the position of  the \CHANDRA source
W46 (X9 in VH98) with the one of its putative counterpart UVE-3 (V1 in
PDF92).  The  choice of  the V1/X9 pair  as a {\it  fiducial position
reference} is justified by the work of many authors (see, for example,
Geffer,  Auriere \& Koch-Miramond  1997 and  VH98), which  confirm the
physical association  between the two objects.  The  shift between the
two  coordinate systems  was admittedly  small: $\delta  X=1''$ ($\sim
20$~ \PC  pixels) and  $\delta Y =0\farcs1$  ($\sim 2$ \PC  ~pixels). \\
After registering the objects  coordinates on a unique reference grid,
we cross correlated  the positions of the X-ray  sources with the ones
derived here for the blue objects.  In this way we found an acceptable
($d<1\farcs5$) positional  coincidence for at least  25 X-ray sources.
The  identification of all  the possible  optical counterparts  to the
X-ray sources  of G01 are listed  in Table~5 while  their positions in
the UV CMD are marked in Figure  4.  Since we intend to use Table 5 as
a  {\it  working progress  list}  and  not  a definitive  counterparts
catalog,  in   some  cases  we  listed  two   UV-objects  as  possible
counterparts to a  single X-ray source (see the case  of W28, W55, W80
and  W98). \\  It is  interesting to  note that  at least  12 possible
optical counterpart  to the X-ray  sources are {\it  faint-UVE} stars,
suggesting, once  more, the existence  of a large population  of faint
IBs  in the  core  of 47  Tuc.   We confirm  the possible  association
(proposed by  VH98) between V2 and X19  (W30 in the G01  list), and we
propose (in agreement with G01) the  association of V3 and X10 (W27 in
the G01 list). Other  position-wise associations appears possible from
our  data-set (see Table~5 for the complete list).
 In  particular, these  are the  ones between  {\it (i)}
UVE-2 (the bright blue  object discussed in (\S \ref{subsec:uve})) and
the X-ray source W75, {\it (ii)} AKO9 (BSS26 in our list - see Section
3.1) and source W36,  {\it (iii)} UVE-6 and UVE-9  and sources W34 and
W105, respectively. \\
However, it  should be  noted that all  the positions of  the putative
 optical   counterparts   could   be   still  affected   by   residual
 uncertanties,  since  the shifts  applied  to  match  the V1/X9  pair
 represent only a first-order  correction. Even though, establishing a
 physical connection  supported only on the  positional coincidence is
 risky,  expecially because of  the large  number of  peculiar objects
 found in the  crowded core of 47~Tuc.  In view  of all these caveats,
 we conservatively conclude that the identifications proposed here are
 possible, and as such useful reference for future investigations, but
 not conclusive.

\subsection{MSPs} 
\label{subsec:msp}

Freire et al. (2001) have recently published accurate timing positions
(typical  uncertainty $<0\farcs1$)  for  15  MSPs in  the  core of  47
Tuc. Eight of these pulsars have  been found in binary systems and, of
these, 3  (namely MSP-I, MSP-O,  MSP-T) lie in  the \PC field  of view.
Exchange interactions  in the core  of the cluster could  in principle
produce a  significant number of systems  comprising a MSP  and a main
sequence  star.  If  the MSP  is an  active X-ray  source,  its energy
emission  could  also affect  such  not-degenerate companion,  perhaps
making it a blue source.   For this reason we cross-correlated the MSP
positions with our  master source list, and we  have found interesting
matches both with the \CHANDRA  X-ray source and the UV object lists.
First,  we confirmed  the finding  by G01  that three  \CHANDRA sources
(namely W19,  W105, W39) are  nearly coincident with the  three binary
MSP (MSP-I, MSP-T, MSP-O, respectively) lying in the \PC field of view.
In addition, we found  interesting matches with peculiar blue objects.
In particular,  MSP-I is at $1\farcs2$  from a white  dwarf (WD-4) and
MSP-T  is  nearly  coincident   (at  only  $0\farcs3$)  with  a  faint
ultraviolet  excess  object (UVE-9).   The  case  of  $MSP-O$ is  more
complex. Only a TO star (namely $\#2167$) has been found at $0\farcs4$
from the nominal position of  this $MSP$.  However, $MSP-O$ is located
in the  region of  the peculiar  binary AKO9 (BSS26  in our  list, see
Section 3.1). Unfortunately this objects is relatively far away ($\sim
2''$)  from  the  nominal  position  of $MSP-O$  to  be  considered  a
promising  optical counterpart  candidate to  this MSP.  Moreover, the
matching with the  X-ray \CHANDRA catalog suggests that  AKO9 is nearly
coincident with another \CHANDRA  source: W36 (see Table~5).  Thus, the
physical  association between  $MSP-O$  and the  peculiar binary  AKO9
appears unlikely.  On the other hand, the only other blue object found
in its  vicinity (at  $\sim 1\farcs5$) is  AKO6 (which  corresponds to
BSS2 in our  list). This object did not  show significant variability,
since it  is not coincident with  any of the 13  variables detected by
E96.  \\ The case of a  possible association between MSP-I and WD-4 is
probably the  more intriguing since  MSP companions in  binary systems
are indeed  expected to be WDs  (e.g. see the reviews  of Verbunt 1993
and Phinney \& Kulkarni 1994),  with cooling ages longer than at least
a few $10^8$ yrs (Hansen  \& Phinney 1998).  According to its position
along the WD cooling sequence, WD-4  seems to be a quite young object,
approximately  $10^6$ yr  old.  It  is thus  unlikely that  it  is the
companion to MSP-I.  Although the irradiation of the surface layers of
the WD by the MSP beam could modify its photometric properties, making
it brighter and slightly bluer  (e.g.  D'Antona 1995 for a review) and
eventually producing  an apparent rejuvanation, this  would require an
unlikely variation of  4-5 magnitudes to reconcile the  cooling age of
$WD-4$  with the  expected cooling  age for  a typical  MSP companion.
Moreover, the induced brightening of  the WD should be recognizable by
a  strong modulation  in its  light  curve, while  its heated  surface
rotates in  and out of  the line of  sight (see e.g.  Stappers  et al.
1999).  Conversely, from the analysis  of the F218W images we have not
found any evidence  for luminosity variation larger than  a few tenths
of magnitude for $WD-4$.  However,  it should be noted that the images
analyzed  here do  not allow  a proper  time coverage  of  the orbital
period of the MSP-I system. In  fact, the F218W images cover less than
$\sim 16\%$  of the  $P\sim 5.5  h$ period estimated  by Freire  et al
(2001).  For  this reason,  we  cannot  exclude  variability for  this
system.  \\ Finally,  UVE-9 is  closely blended  with a  main sequence
star,  and its  photometry turns  presently  to be  too uncertain  for
allowing further speculations. \\ Table~6 lists the absolute positions
for the three MSPs and for  the blue sources lying nearby. However, in
none  of  the above  cases  the  positional  coincidence alone  is  an
argument  strong  enough   to  support  the  identification.   Indeed,
astrophysical and  statistical considerations suggest that  all of the
above  positional  coincidences  are  probably due  to  chance,  thus,
probably the optical  counterparts to these MSPs are  much fainter than
the  detection  limit  of  the  observations  analyzed  here.  \\  The
probabilities of a physical connection between these 3 binary MSP systems
and the  blue objects  in their proximity  are further reduced  if one
examines also the optical sources  located in the vicinity of the {\it
single, isolated,} MSPs detected in the core of 47~Tuc and included in
the field of view covered  by the PC.  Although no optical counterpart
is expected  for these objects, it  appears (see Table~6)  that two of
them (over  the three non-binary MSP  listed by Freire  et al.  2001),
namely MSP-F and MSP-L,  are positionally coincident, within the above
uncertainties, with WD-8 and  BSS-29, respectively.  \\ While reaching
fainter  limiting magnitudes would  certainly help  the search  to MSP
companions, the  current uncertainties  on the absolute  \PC astrometry
would  still represent  the major  problem, suggesting  that  even the
availability of much deeper exposures would not solve the ambiguities.
We can  use Eq.   (1) in order  to compute  the expected number  of WD
(younger  than a  given cooling  time) lying  in the  fraction  of the
47~Tuc core covered by the PC.  In doing this, we assumed a WD cooling
age of 1  Gyr, i.e.  at the median of the  WD cooling age distribution
of  $\sim 10^8-10^{10}$  yrs  as suggested  by  some $WD+MSP$  systems
studied  in the  field  (see Hansen  \&  Phinney 1998;  Sch\"onberner,
Driebe  \& Bl\"ocker  2000).  Adopting  the Wood  (1995)  models, this
cooling age  corresponds to an absolute magnitude  $M_V\sim 13$ (which
yealds $m_{F218W}\sim  27$ at  the distance of  47 Tuc).   Under these
assumptions,  Eq.   (1)  gives  a   total  of  $\sim  1000$  WDs  with
$m_{F218W}\lsim  27$. Assuming that  $\sim 2$  \arcsec is  the typical
uncertainty in  registering the MSP  coordinates on the \PC  frame, the
previous   number  implies  an   average  of   6  WDs   brighter  than
$m_{F218W}=27$  within  the  uncertainty  region.  The  situation  can
certainly improved by obtaining a  new astrometric solution for the PC
using the GSC2.2  (McLean et al. 2001), which  has an average absolute
astrometric accuracy of $\sim  0\farcs5$ i.e.  three times better than
the one  of the GSC1.1.  However,  even a remarkable  improvement of a
factor  of 3  in the  absolute astrometry  of the  \PC will  reduce the
probability of  a chance coincidence between  the MSP and  a 1 Gyr-old
WD, only  to a $\sim  70\%$, still too  high to claim  a position-wise
identification. \\ In summary, the conspiracy of the overcrowded field
in  the core  of  47~Tuc and  the  extreme faintness  of the  expected
optical  counterpart (an  old  WD) of  the  companion to  a MSP  makes
extremely difficult a firm identification using only photometric colors
and  positioning.   Further  hints  will  be necessary,  such  as  the
discovery   of  photometric   modulations  due   to   orbital  motion,
peculiarities in the optical and UV companion candidate spectra and/or
emissivity at other wavelengths.
 
\section{Conclusions}
\label{sec:con}

Using data  collected with the Planetary  Camera of HST,  we have more
than doubled the sample of Blue Stragglers detected in the core of the
globular cluster 47~Tuc and have  reported on the discovery of a large
number of UV  stars, with the hottest of them being probably young WDs.  The  others UV objects  are {\it anomalous}
UV-excess   stars  whose   true   nature  and   evolution  cannot   be
satifactorily  assessed  yet.   In  fact many  different  evolutionary
mechanisms involving stellar collisions and interactions could account
for   them.  
High  resolution,   deeper  imaging   and  spectroscopic
observations in  the UV (as  those preliminary presented by  Knigge et
al. 2001) are required to  discriminate among the various models.
However, the large number of positional coincidences
with X-ray sources (from the recent \CHANDRA  catalog)
 indeed suggests that part of 
them could be IBs. In particular, a significant number of X-ray sources
($\sim 30\%$) have been
found to be possibly associated with the {\it faint UVE} population
discovered here, supporting the presence of a large population of
faint CV in the core of this cluster.\\ We
have also explored the possibility of identifying some of the detected
blue  objects  as  the  optical  counterparts  to 
millisecond  pulsar companions  hosted  in the  core  of the  cluster.
However, although a few positional coincidences
 have been found, none of them is 
suggestive yet of a physical association.

\acknowledgements

The financial  support of the  Agenzia Spaziale Italiana (ASI)  and of
the {\it  Ministero della Universit\`a  e della Ricerca  Scientifica e
Tecnologica} (MURST) to the  project {\it Stellar Dynamics and Stellar
Evolution in Globular Clusters} is kindly acknowledged.

\clearpage
 
\begin{deluxetable}{cccccc}
\scriptsize
\tablewidth{12cm}
\label{tab:uve}
\tablecaption{Description of the used data-sets. The first two 
columns give the date 
of the observation and the proposal ID number. 
The filter name, its pivot wavelength and width are given in 
the second and 
third columns, respectively. The number of exposures and the integration time 
per exposure are given in the last two columns.}
\tablehead{
\colhead{Date} &
\colhead{Proposal ID} &
\colhead{Filter} &
\colhead{$\lambda$/$\Delta\lambda$} & 
\colhead{N.exp} &
\colhead{Exp (s)}   
}
\startdata
1995-09-01 & GO-6095 & F218W & 2189  & 4 &    800 \\
1999-07-11 & GO-8267 & F336W & 3341  & 6 &    900 \\
1995-09-01 & GO-6095 & F439W & 4300  & 2 &    50  \\
\enddata
\end{deluxetable}

\clearpage 

\begin{deluxetable}{ccccccccc}
\scriptsize
\tablewidth{15cm}
\label{tab:bss}
\tablecaption{BSS candidates in 47~Tuc. The columns give the 
BSS-\# according to the notation of P91, the  object ID 
number, the magnitudes
in the three filters and the  object coordinates in \PC pixels, as measured 
relative to the F218W 
image reference frame. The last column gives the identifications found in the 
literature.}
\tablehead{
&
 \colhead{BSS \#} &
 \colhead{ID} &
\colhead{$m_{F218W}$} & 
\colhead{$m_{F336W}$} &
\colhead{$m_{F439W}$} & 
\colhead{X} &
\colhead{Y} & 
\colhead{G92/E96}  
 }
\startdata
& BSS-1  &   1893 & 15.92 & 15.34 & 14.46 &  556.60 &  400.85 & 172 \\
& BSS-2 &   2314  &    17.63 & 16.29 & 15.72 &  498.78 &  325.23 & 206(AKO6)\\
& BSS-3 &   3084  &    17.99 & 16.76 & 16.21 &  438.14 &  178.80 &  -- \\
& BSS-4 &    997  &   17.20 & 16.38 & 15.78 &  421.77 &  556.46 & 302 \\
& BSS-5 &   2094  &    16.76 & 16.08 & 15.26 &  392.19 &  365.54 & 299(V6)\\
& BSS-6 &   3132  &    17.86 & 16.87 & 16.30 &  358.08 &  171.09 &  -- \\
& BSS-7 &   3255  &   17.15 & 16.32 & 15.68 &  348.52 &  145.64 & 312 \\
& BSS-8 &   2004  &   17.79 & 16.72 & 16.12 &  376.00 &  382.06 &  -- \\
& BSS-10 &   1420  &    17.73 & 16.62 & 16.06 &  367.75 &  480.76 & -- \\
& BSS-12 &   1654  &    18.69 & 17.00 & 16.49 &  351.66 &  442.10 & -- \\
 &BSS-13  &   2025  &    17.01 & 16.06 & 15.25 &  307.85 &  378.44 & 386\\
& BSS-14 &   1835  &   18.19 & 16.92 & 16.33 &  309.64 &  409.05 &  -- \\
 &BSS-15 &    694  &    17.12 & 16.29 & 15.65 &  344.68 &  622.69 & 389\\
& BSS-16 &    947  &    17.28 & 16.50 & 15.89 &  332.21 &  565.74 & V12\\
& BSS-18 &   1561  &     16.92 & 16.20 & 15.47 &  308.13 &  459.84 & 398(V3)\\
& BSS-19 &   1764  &    16.38 & 15.82 & 14.87 &  157.26 &  422.67 & 534\\
& BSS-20  &   2556  &   18.04 & 16.77 & 16.26 &  110.84 &  281.95 &  --\\
& BSS-21 &    642  &  18.45 & 16.90 & 16.41 &  211.47 &  632.92 &  --\\
& BSS-22 &    183  &    17.06 &   --  & 15.24 &  373.60 &  742.95 & 381(V1)\\
& BSS-23 &   3086  &    15.96 & 15.54 & 14.48 &  262.24 &  178.40 & 399 \\ 
& BSS-24&   2835  &    18.07 & 17.05 & 16.45 &  127.40 &  227.41 & V10\\ 
& BSS-25 &   2675  &    16.46 & 15.74 & 14.85 &  712.99 &  257.58 & 32 \\ 
& BSS-26  &   2324  &   19.21 & 17.64 & 17.52 &  451.66 &  323.77 &V11(AKO9)\\
\enddata
\end{deluxetable}

\addtocounter{table}{-1}
\begin{deluxetable}{ccccccccc}
\scriptsize
\tablewidth{15cm}
\tablecaption{(continue)}
\tablehead{&
\colhead{BSS \#} &\colhead{ID} &
\colhead{$m_{F218W}$} & 
\colhead{$m_{F336W}$} &
\colhead{$m_{F439W}$} & 
\colhead{X} &
\colhead{Y} & 
\colhead{G92/E96}  
 }
\startdata
& BSS-27 &   3212  &    19.22 & 17.85 & 17.30 &  255.84 &  153.44 &  -- \\  
& BSS-28 &   2947  &    18.56 & 16.65 & 16.05 &  251.63 &  204.62 &  -- \\
& BSS-29&   2712  &   17.63 & 16.79 & 16.23 &  529.18 &  250.89 &  -- \\
& BSS-30 &   2637  &    19.02 & 17.41 & 16.85 &  122.77 &  265.71 &  -- \\
& BSS-31 &   2364  &    17.90 & 16.00 & 15.45 &  645.27 &  315.88 &  -- \\
& BSS-32 &   1901  &    18.51 & 17.30 & 16.78 &  471.27 &  399.25 &  -- \\
& BSS-33 &   1838  &    18.10 & 16.57 & 16.01 &  574.08 &  408.72 &  -- \\
 &BSS-34 &   1819  &    19.26 & 17.52 & 17.04 &  296.37 &  411.62 &  -- \\
& BSS-35 &   1603  &    19.34 & 17.10 & 16.63 &  327.81 &  451.06 &  -- \\
& BSS-36 &   1393  &    18.94 & 17.61 & 17.17 &  310.02 &  484.39 &  -- \\
 &BSS-37 &   2121  &   18.79 &   --  & 16.85 &  741.28 &  360.81 &  -- \\
& BSS-38 &   1865  &   19.07 & 17.89 & 17.25 &  464.72 &  404.88 &  -- \\
& BSS-39 &   1746  &   18.71 &   -- & 16.33 &  642.58 &  425.26 &   -- \\
 &BSS-40 &    568  &   17.37 &   -- & 15.73 &  460.89 &  644.52 &   -- \\
& BSS-41 &    561  &   19.22 &   -- & 16.76 &  542.56 &  645.08 &   -- \\
& BSS-42 &    421  &   16.90 &   -- & 15.64 &  149.71 &  680.89 &   -- \\
& BSS-43 &    199  &   19.35 &   -- & 17.44 &   64.80 &  739.58 &   -- \\ 
 &BSS-44 &    157  &   17.17 &   -- & 15.61 &  734.10 &  753.10 &   -- \\
& BSS-45 &     53  &   18.44 &   -- & 16.73 &  324.11 &  780.87 &   -- \\
& BSS-46 &   2324  &   19.21 & 17.64 & 17.52 &  451.66 &  323.77 &  -- \\
\enddata
\end{deluxetable}

\clearpage

\begin{deluxetable}{cccccccc}
\scriptsize
\tablewidth{15cm}
\label{tab:wd}
\tablecaption{Young WD candidates in 47~Tuc. Columns are as in Table 2.}
 \tablehead{&\colhead{WD \#} &
\colhead{ID} &
\colhead{$m_{F218W}$} & 
\colhead{$m_{F336W}$} &
\colhead{$m_{F439W}$} & 
\colhead{X} &
\colhead{Y}   
 }
\startdata
 & WD-1   & 3561   &   17.41 &  -- & 19.56 &  780.63 &   78.47 \\
 & WD-2   &  302  &    18.13 & 19.60 & 20.64 &  358.87 &  713.66 \\
 & WD-3   & 626   &    18.70 & 20.01 & 21.32 &  395.86 &  635.04 \\
 & WD-4   &  234  &    19.19 &  -- & 21.35 &  201.56 &  733.14 \\
 & WD-5   &3036   &    19.26 & 20.52 & 21.60 &  364.93 &  186.49 \\
 & WD-6   & 2468  &    19.51 & 20.57 & 21.49 &  357.58 &  299.51 \\
 & WD-7   & 111   &    19.53 &  -- & 21.74 &  304.96 &  765.47 \\
 & WD-8   & 728   &    19.56 & 21.10 & 21.50 &  414.00 &  614.64 \\
  &WD-9   & 945   &    19.80 &  -- & 21.97 &  588.24 &  565.99 \\
 & WD-10  &  3259 &    19.82 & 20.84 & 21.95 &  427.11 &  144.16 \\
 & WD-11  & 2716  &   19.97 & 21.44 & 22.25 &  404.79 &  250.14 \\
 & WD-12  & 3251  &    20.06 & 21.26 & 21.93 &  687.62 &  146.26 \\

\enddata
\end{deluxetable}
\clearpage

\begin{deluxetable}{cccccccc}
\scriptsize
\tablewidth{15cm}
\label{tab:uve}
\tablecaption{Bright UV-Excess stars (UVEs) in 47~Tuc.Columns are as in Table 2 
and 3.}
\tablehead{&\colhead{UVE \#} &
\colhead{ID} &
\colhead{$m_{F218W}$} & 
\colhead{$m_{F336W}$} &
\colhead{$m_{F439W}$} & 
\colhead{X} &
\colhead{Y}   
 }
\startdata
 & UVE-1 &  247   &   17.34 &  --  & 17.96 &  290.15 &  730.87 \\
  &UVE-2 & 1798   &   18.05 & 18.03 & 17.16 &  303.81 &  416.16 \\
  &UVE-3 & 2676   &    18.52 & 18.89 & 19.61 &  510.36 &  256.90 \\
  &UVE-4 &1414   &   18.77 & 19.77 & 20.33 &  327.73 &  481.32 \\
 & UVE-5 &1406   &   19.35 & 20.09 & 20.25 &  110.73 &  483.36 \\
  & UVE-6 &1046   &  19.50 & 20.93 & 20.81 &  454.11 &  546.51 \\
  &UVE-7 & 3462   &  19.63 & 20.92 & 20.15 &  156.73 &   98.93 \\
  & UVE-8 &3378   &  19.95 & 19.98 & 19.13 &  667.83 &  119.54 \\ 
  & UVE-9 &225    &  20.13 &  --  & 19.64  &  122.93  &  735.11 \\ 
 & UVE-10 & 2388   & 20.22 & 21.65 & -- &  682.81 &  310.62 \\
  & UVE-11 &1907   &  20.26 & 20.50 & 19.97 &  263.85 &  398.23 \\
\enddata
\end{deluxetable}

\clearpage

\begin{deluxetable}{ccccccccc}
\scriptsize
\tablewidth{17cm}
\label{tab:xray}
\tablecaption{\CHANDRA X-ray sources and associated blue objects. The last 
two columns give the nominal angular separation and a classification/name 
for the X-ray  source or the UV object.}
 \tablehead{
\colhead{X-Name}&
\colhead{Name}&
\colhead{$m_{F218W}$} & 
\colhead{$m_{F336W}$} &
\colhead{$m_{F439W}$} & 
\colhead{$X$ }&
\colhead{$Y$ }&
\colhead{$dist$ }&
\colhead{Other Names}
}
\startdata
 W15 & 2452  & 20.72 &  --  & 20.23 &   69.01 &  302.93 &   1\farcs 4 & 
$faint-UVE$\nl
 W19 &  234  & 19.19 &  --  & 21.35 &  201.56 &  733.14 &   0\farcs 9 & WD4\nl
 W24 & 1176  & 20.56 &  --  & 19.14 &  236.01 &  521.26 &   1\farcs 4 & 
$faint-UVE$\nl
 W27 &  716  & 20.66 & 21.55 &  --   &  346.74 &  617.20 &   0\farcs 0 
&X10,V3\nl
 W28 & 2800  & 18.09 & 15.58 & 14.80 &  335.29 &  233.99 &   1\farcs 5 & BSS9 
\nl
 W28 & 3066  & 21.17 &   -- & 18.92 &  334.03 &  181.80 &    1\farcs 2 & 
$faint-UVE$\nl
 W29 & 1408  & 19.55 & 18.04 & 17.55 &  361.41 &  482.70 &   0\farcs 1 & 
$faint-BSS$\nl
 W30 & 5400  & 20.70 & 19.41 & 20.21 &  340.85 &  325.13 &   0\farcs 1 & 
X19,V2\nl
 W31 & 3255  & 17.15 & 16.32 & 15.68 &  348.52 &  145.64 &   0\farcs 4 & BSS7\nl
 W34 & 1046  & 19.50 & 20.93 & 20.81 &  454.11 &  546.51 &   1\farcs 3 & 
UVE-6\nl
 W36 & 2324  & 19.21 & 17.64 & 17.52 &  451.66 &  323.77 &   0\farcs 1 
&BSS26,AKO9,V11\nl
 W37 & 1865  & 19.07 & 17.89 & 17.25 &  464.72 &  404.88 &   0\farcs 5 &BSS38 
\nl
 W39 & 2314  & 17.63 & 16.29 & 15.72 &  498.78 &  325.23 &   1\farcs 5 & 
BSS2,AKO6\nl
 W42 & 2676  & 18.52 & 18.89 & 19.61 &  510.36 &  256.90 &   0\farcs 0 & 
UVE3,X9,V1\nl
 W44 & 2907  & 21.58 &  --  & 20.00 &  531.93 &  212.76 &   1\farcs 5 & 
$faint-UVE$\nl
 W47 & 3611  & 21.54 &  --  & 19.77 &  558.48 &   68.68 &   0\farcs 9 & 
$faint-UVE$\nl
 W49 & 1441  & 20.68 &  --  & 18.79 &  694.42 &  477.85 &   1\farcs 4 & 
$faint-UVE$\nl
 \enddata
\end{deluxetable}
 
\clearpage
\addtocounter{table}{-1}
\begin{deluxetable}{ccccccccc}
\scriptsize
\tablewidth{17cm}
\label{tab:xray}
\tablecaption{(continue)}
 \tablehead{
\colhead{X-Name}&
\colhead{Name}&
\colhead{$m_{F218W}$} & 
\colhead{$m_{F336W}$} &
\colhead{$m_{F439W}$} & 
\colhead{$X$ }&
\colhead{$Y$ }&
\colhead{$dist$ }&
\colhead{Other Names}
}
\startdata
 W51 & 1646  & 21.08 &  --  & 19.48 &  678.11 &  443.09 &   0\farcs 8 & 
$faint-UVE$\nl
 W54 & 1146  & 19.36 &  --  & 15.50 &  733.57 &  526.64 &   0\farcs 1 & \nl
 W55 & 2121  & 18.79 &  --  & 16.85 &  741.28 &  360.81 &   1\farcs 1 & BSS37\nl
 W55 & 2141  & 20.88 &  --  & 18.81 &  721.60 &  358.50 &   1\farcs 5 & 
$faint-UVE$\nl
 W73 & 2307  & 19.97 & 18.85 & 18.33 &  190.41 &  326.10 &   1\farcs 3 
&$faint-UVE$\nl
 W75 & 1798  & 18.05 & 18.03 & 17.16 &  303.81 &  416.16 &   0\farcs 7 
&UVE2,V36\nl
 W77 &  945  & 19.80 &  --  & 21.97 &  588.24 &  565.99 &   0\farcs 9 & WD9\nl
 W80 &  819  & 20.88 &  --  & 19.16 &  728.91 &  595.90 &   0\farcs 2 
&$faint-UVE$ \nl
 W80 &  882  & 20.87 &  --  & 20.49 &  698.91 &  580.20 &   1\farcs 4 
&$faint-UVE$ \nl
 W98 & 3259  & 19.82 & 20.84 & 21.95 &  427.11 &  144.16 &   1\farcs 1 & WD10 
\nl
 W98 & 3099  & 21.45 & 20.58 & 19.17 &  423.93 &  176.73 &   0\farcs 7 
&$faint-UVE$ \nl
 W105 &  225  & 20.13 &  -- & 19.64 &  122.93 &  735.11 &   1\farcs 0 &UVE9\nl
 \enddata
\end{deluxetable}

\clearpage
 
\begin{deluxetable}{ccccc}
\scriptsize
\tablewidth{12cm}
\label{tab:msp}
\tablecaption{MSP and blue object positions in 47~Tuc. The 
 name of the \CHANDRA X-ray source
possibly associated with the MSP 
  is also reported (in parenthesis).}
\tablehead{
&
\colhead{Name} &
\colhead{$\alpha_{2000}$ } &
\colhead{$\delta_{2000}$ }& 
\colhead{d}  
}
\startdata
&$ MSP-I(W19)$ & $ 00^{\rm h}\, 24^{\rm m}\, 07\fs 93$ & $-72\arcdeg\, 
04\arcmin\, 
39\farcs 66$ & $--$\\
& $WD-4$  & $ 00^{\rm h}\, 24^{\rm m}\, 07\fs 97$ & $-72\arcdeg\, 04\arcmin\, 
38\farcs 80$ & $0\farcs 8$\\ 
\tableline
&$ MSP-T(W105)$ & $ 00^{\rm h}\, 24^{\rm m}\, 08\fs 55$ & $-72\arcdeg\, 
04\arcmin\, 
38\farcs 90$ & $--$\\
& $UVE-9$& $ 00^{\rm h}\, 24^{\rm m}\, 08\fs 73$ & $-72\arcdeg\, 04\arcmin\, 
39\farcs 24$ & $0\farcs 9$\\ 
\tableline
&$ MSP-O(W39)$ & $ 00^{\rm h}\, 24^{\rm m}\, 04\fs 65$ & $-72\arcdeg\, 
04\arcmin\, 
53\farcs 75$ & $--$\\
& $BSS-2$& $ 00^{\rm h}\, 24^{\rm m}\, 04\fs 48$ & $-72\arcdeg\, 
04\arcmin\, 54\farcs 99$
 & $1\farcs 5$\\ 
\tableline
\tableline
&$ MSP-L$ & $ 00^{\rm h}\, 24^{\rm m}\, 03\fs 77$ & $-72\arcdeg\, 04\arcmin\, 
56\farcs 91$ & $--$\\
& $BSS-29$&  $ 00^{\rm h}\, 24^{\rm m}\, 04\fs 07$ & $-72\arcdeg\, 04\arcmin\, 
58\farcs 10$ & $1\farcs 7$\\ 
\tableline
&$ MSP-F(W77)$ & $ 00^{\rm h}\, 24^{\rm m}\, 03\fs 85$ & $-72\arcdeg\, 
04\arcmin\, 
42\farcs 80$ & $--$\\
& $WD-9$&  $ 00^{\rm h}\, 24^{\rm m}\, 03\fs 97 $ & $-72\arcdeg\, 04\arcmin\, 
43\farcs 56$ 
& $0\farcs 9$\\ 
\enddata
\end{deluxetable}

\clearpage

\begin{figure}
\plotone{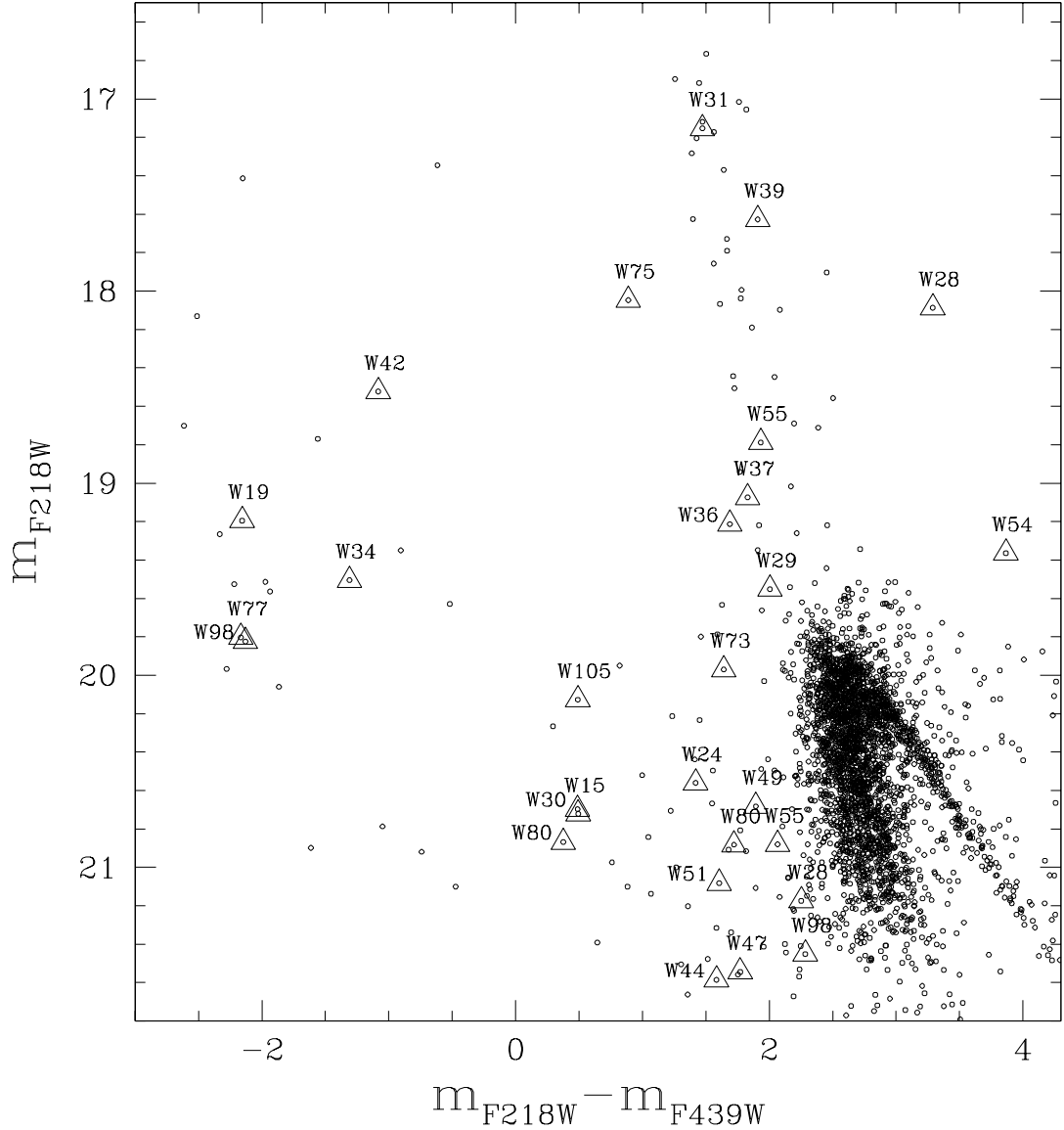} 
\caption{Zoom   of  the   ($m_{F218W},   m_{F218W}-m_{F439W}$)-CMD  of
Figure~\ref{fig:map439}.
The optical counterpart candidates to the \CHANDRA
X-ray source are marked with large open triangles and
labelled with their names (from Table 1 by G01).
\label{fig:x}}
\end{figure}

\end{document}